\documentclass[fleqn,12pt]{article}
\usepackage{amssymb,latexsym}
\usepackage[OT2,OT1]{fontenc}

\usepackage{url}
\usepackage{hyperref}
\input xy
\xyoption{all}
\DeclareFontFamily{OT1}{rsfs}{}
\DeclareFontShape{OT1}{rsfs}{m}{n}{ <-7> rsfs5 <7-10> rsfs7 <10->
rsfs10}{} \DeclareMathAlphabet{\mycal}{OT1}{rsfs}{m}{n}

\begin{document}
\title{\bf Conformal Yano-Killing tensors for the Taub-NUT metric}
\author{Jacek Jezierski\thanks{Partially supported by
    EPSRC: EP/D032091/1. E--mail: \texttt{Jacek.Jezierski@fuw.edu.pl}}
     \ and Maciej {\L}ukasik \\ Department of Mathematical Methods in
Physics, \\ University of Warsaw, ul. Ho\.za 69, 00-681 Warsaw,
Poland}
\maketitle
{\catcode `\@=11 \global\let\AddToReset=\@addtoreset}
\AddToReset{equation}{section}
\renewcommand{\theequation}{\thesection.\arabic{equation}}

\newtheorem{Definition}{Definition}
\newtheorem{Lemma}{Lemma}
\newtheorem{Theorem}{Theorem}
\newtheorem{Remark}{Remark}
\newtheorem{Proposition}{Proposition}
\newcommand{\gn}[2]{\stackrel{\mbox{\tiny (#1)}}{#2}}
\newcommand{\gthree}{h}
\newcommand{\TBR}{T^{\scriptscriptstyle BR}}
\newcommand{\TEM}{T^{\scriptscriptstyle EM}}
\newcommand{\QBR}{CQ^{\scriptscriptstyle BR}}
\newcommand{\QEM}{CQ^{\scriptscriptstyle EM}}
\newcommand{\QYK}{\Theta}
\newcommand{\Img}{{\rm Image}}
\newcommand{\dtwo}{\mathbf{\Delta}}
\newcommand{\kolo}[1]{\vphantom{#1}\stackrel{\circ}{#1}\!\vphantom{#1}}
\newcommand{\eq}[1]{(\ref{#1})}
\newcommand{\arxiv}[1]{\url{http://arxiv.org/abs/#1}}
\newcommand{\rd}{\,{\rm d}} 
\newcommand{\tr}{\mathop {\rm tr}\nolimits }
\newcommand{\trg}{\mathop {\rm tr_g}\nolimits }
\newcommand{\arsh}{\mathop {\rm arsh}\nolimits }
\newcommand{\be}{\begin{equation}}
\newcommand{\ee}{\end{equation}}
\newcommand{\ber}{\begin{eqnarray}}
\newcommand{\eer}{\end{eqnarray}}
\newcommand{\tg}{{\tilde g}}
\newcommand{\tGam}{{\tilde\Gamma}}
\newcommand{\tD}{{\tilde D}}
\newcommand{\tL}{{\tilde L}}
\newcommand{\N}{{\mathbb N}}
\newcommand{\tR}{{\tilde R}}
\newcommand{\mcY}{{\cal Y}}
\newcommand{\ped}{{\chi}}
\newcommand{\Exp}{{\rm e}}


\newcounter{mnotecount}[section]
\renewcommand{\themnotecount}{\thesection.\arabic{mnotecount}}
\newcommand{\mnote}[1]
{\protect{\stepcounter{mnotecount}}$^{\mbox{\footnotesize  $
      \bullet$\themnotecount}}$ \marginpar{\raggedright\tiny
    $\!\!\!\!\!\!\,\bullet$\themnotecount: #1} }
\newcommand{\JJ}[1]{\mnote{\textbf{JJ:} #1}}

\begin{abstract}
Symmetric conformal Killing tensors and (skew-symmetric) conformal
Yano--Killing tensors for Euclidean Taub-NUT
metric are given in explicit form.  Relations between Yano and CYK tensors
in terms of conformal rescaling are discussed.
\end{abstract}

\section{Introduction}
In \cite{JJMLKerr} we examined conformal Yano--Killing tensors in Kerr
spacetime. In this paper we discuss Euclidean Taub-NUT metric which is also an interesting
case possessing non-trivial CYK tensors.

According to \cite{cykem} one can define, in terms of spacetime
curvature, two kinds of conserved quantities with the help of
conformal Yano--Killing tensors (see \cite{Tachibana}, \cite{Yano}).
Sometimes they are also called conformal
Killing forms or twistor forms (see e.g. \cite{Moroianu},
\cite{Semmelmann}, \cite{Stepanow}). The first kind is linear and
the second quadratic with respect to the Weyl tensor but a basis
for both of them is the Maxwell field. Conserved quantities which
are linear with respect to CYK tensor were investigated many times
(cf. 
\cite{Glass-Naber}, 
\cite{Goldberg1}, 
\cite{JJspin2}, \cite{kerrnut}, \cite{cykem}, \cite{Pen1}, 
\cite{Pen-Rin}). On the other hand, quadratic charges are less
known and have usually been examined in terms of the Bel--Robinson
tensor (see e.g. \cite{Berg}, \cite{Sen}, \cite{Ch-Kl},
\cite{Douglas}).

This paper is organized as follows: In the next Section we introduce
basic notions, CYK tensors for Euclidean Taub-NUT metric
and derive conformal symmetric Killing tensors.
In Section 3 we analyze
the question \emph{if we can reduce Conformal Yano-Killing tensor
to Yano by conformal transformation?}

\section{Taub-NUT metrics and its CYK tensors}\label{Kerr_Taub-NUT_ch}
Let $M$ be an $n$-dimensional ($n>1$) manifold with a Riemannian
or pseudo-Riemannian metric $g_{\mu\nu}$. The covariant derivative
associated with the~Levi--Civita connection will be denoted by
$\nabla$ or just by ``$\,;\,$''. By $T_{...(\mu\nu)...}$ we will
denote the symmetric part and by $T_{...[\mu\nu]...}$ the
skew-symmetric part of tensor $T_{...\mu\nu...}$ with respect to
indices $\mu$ and $\nu$ (analogous symbols will be used for more
indices).

Let $Q_{\mu\nu}$ be a skew-symmetric tensor field (two-form) on
$M$ and let us denote by ${\cal Q}_{\lambda \kappa \sigma}$ a
(three-index) tensor which is defined as follows:
\begin{equation}\label{CYK_eq1}
    {\cal Q}_{\lambda \kappa
    \sigma}(Q,g):= Q_{\lambda \kappa ;\sigma} +Q_{\sigma \kappa
    ;\lambda} - \frac{2}{n-1} \left( g_{\sigma
    \lambda}Q^{\nu}{_{\kappa ;\nu}} + g_{\kappa (\lambda }
    Q_{\sigma)}{^{\mu}}{_{ ;\mu}} \right) \, .
\end{equation}
The object $\cal Q$ has the following algebraic properties
\be\label{wlQ}
   {\cal Q}_{\lambda\kappa\mu}g^{\lambda\mu}=0=
   {\cal Q}_{\lambda\kappa\mu}g^{\lambda\kappa} \, , \quad
{\cal Q}_{\lambda\kappa\mu} = {\cal Q}_{\mu\kappa\lambda}\, ,
   \ee
i.e. it is traceless and partially symmetric. In \cite{JJMLKerr}
(see also \cite{JJspin2}, \cite{kerrnut}) we proposed the following
\begin{Definition}\label{CYK_df}
    A skew-symmetric tensor $Q_{\mu\nu}$ is a conformal Yano--Killing tensor
    (or simply CYK tensor) for the metric $g$ iff
    \mbox{${\cal Q}_{\lambda \kappa \sigma}(Q,g) = 0$}.
\end{Definition}
In other words, $Q_{\mu\nu}$ is a conformal Yano--Killing tensor
if it fulfils the following equation:
\begin{equation}\label{CYK_eq2}
    Q_{\lambda \kappa ;\sigma} +Q_{\sigma \kappa ;\lambda} =
    \frac{2}{n-1} \left( g_{\sigma \lambda}Q^{\nu}{_{\kappa ;\nu}} +
    g_{\kappa (\lambda } Q_{\sigma)}{^{\mu}}{_{ ;\mu}} \right) \,
\end{equation}
(first proposed by Tachibana and Kashiwada, cf. \cite{Tachibana}).
Moreover, if $\xi_\mu:= Q^{\nu}{_{\mu ;\nu}}$ vanishes then $Q$ is a usual Yano tensor
i.e. a solution of equation (\ref{CYK_eq2}) with vanishing
right-hand side.

Let us consider Euclidean Taub-NUT metric which is an example of a
metric admitting nontrivial solutions of the
equation~(\ref{CYK_eq2}).  We will define it in terms of
coordinate system $(\psi, r, \theta, \phi)$. In these coordinates
the metric tensor has a form:
\begin{equation}\label{Taub-NUT_metric}
g=\left(1+\frac{2m}{r}\right)\left(\rd
r^2+r^2\rd\theta^2+r^2\sin^2\theta \rd\phi^2\right) +
\frac{4m^2}{1+\frac{2m}{r}}\left(\rd\psi+\cos\theta
\rd\phi\right)^2.
\end{equation}
Passing to the limit as $r\to\infty$ we get:
\begin{equation}\label{Taub-NUT_metric_limit}
g=\rd r^2+r^2\rd\theta^2+r^2\sin^2\theta\rd\phi^2 +
\rd\bar{\psi}^2,
\end{equation}
where $\bar{\psi}:=2m\psi$ (the term $\cos\theta\rd\phi$ is
negligible relative to $r\sin\theta\rd\phi$). It means that for
large $r$ the metric $g$ looks like flat Euclidean metric and
therefore may be called ``asymptotically flat Euclidean metric''.
However,  Taub-NUT manifold (denoted by $M$) is a
bundle $M\to S^2$ with base coordinates $(\theta,\phi)$ on the two-sphere $S^2$. Fibre
coordinates are $(\psi,r)$. Moreover, it is not a trivial bundle,
i.e. it has no global section. If, for example, we restrict
ourselves to points with constant $\psi$ and $r$, than the metric
induced on this surface is singular. We will see that it causes
some difficulties.

Similarly to the case of Kerr metric (cf. \cite{JJMLKerr})
there are known Yano tensors for the
metric (\ref{Taub-NUT_metric}) (see e.g.~\cite{vHolt}).
They are given by the following formulae:
\begin{equation}\label{Taub-NUT_Y}
Y=2m^2\left(\rd\psi+\cos\theta \rd\phi\right) \wedge \rd r
+r\left(r+m\right) \left(r+2m\right) \sin\theta \rd\theta \wedge
\rd\phi,
\end{equation}
\begin{equation}\label{Taub-NUT_Yi}
Y_{i}=4m\left(\rd\psi + \cos\theta \rd\phi\right) \wedge \rd
x_{i}-(1+\frac{2m}{r}) \epsilon_{ijk}\rd x^{j} \wedge \rd x^{k},
\end{equation}
where $1 \le i$, $j$, $k \le 3$. The functions $x^i=x_i$ and the
symbol $\epsilon_{ijk}$ are defined by:
\[
x^1 := r\sin\theta\cos\phi,
\]
\[
x^2 := r\sin\theta\sin\phi,
\]
\[
x^3 := r\cos\theta,
\]

\[
\epsilon_{ijk}=\cases{+1 & if $ijk$ is an even permutation of
1,2,3\cr
                      -1 & if $ijk$ is an odd permutation of  1,2,3\cr
                 \,\ \ 0 & in any other cases}
\]
We can now ask, how tensors dual to $Y$ and $Y_i$ look like. It
turns out that tensors $Y_{i}$ are anti-selfdual, i.e. $\ast Y_{i}
= -Y_{i}$. They are also covariantly constant, i.e.
$\nabla_\rho (Y_i)_{\mu\nu}=0$
Moreover, the two-forms $F_{\mu\nu}(R,Y_i) := R_{\mu\nu\lambda\kappa}(Y_i)^{\lambda\kappa}$,
 where $R$ is a Riemann
tensor\footnote{Ricci tensor of Taub-NUT metric vanishes, which
implies that its Riemann tensor is a spin-2 field.} of the metric
$g$, are identically equal to zero.

Tensor dual to $Y$ has a form:
\begin{equation}\label{Taub-NUT_Ystar}
\ast Y=2m\left(m+r\right)\left(\rd\psi+\cos\theta \rd\phi\right)
\wedge \rd r+mr\left(r+2m \right) \sin\theta \rd\theta \wedge
\rd\phi.
\end{equation}
$*Y$ is not a Yano tensor anymore. Its divergence $\chi$ defined as
$\chi^\nu:=*Y^{\mu\nu}{}_{;\mu}$  equals to
$-\frac{3}{2m}\partial_{\psi}$, which implies that
equation~(\ref{CYK_eq2}) for $*Y$ has nontrivial right-hand side.

We may try to derive ``Euclidean'' charges corresponding to tensors $Y$
and $*Y$ . Asymptotically they look as follows:
\[
Y = r^3\sin\theta \rd\theta \wedge \rd\phi + O\left(1\right) =
*(r \rd\psi \wedge \rd r) + O\left(1\right),
\]
\[
*Y = r\rd\psi \wedge \rd r + O\left(1\right).
\]
We can say, then, that a charge corresponding to $*Y$ is the (Euclidean) energy,
and a charge corresponding to $Y$ is the dual energy (cf. \cite{JJMLKerr}).

It turns out that, although $Y$ and $*Y$ are different tensors,
the forms $F(R,Y)$ and~$F(R,*Y)$ are the same. We will denote it
by $\widetilde{F} := F(R,Y) = F(R,*Y)$. We have:
\begin{equation}\label{Taub-NUT_F}
\widetilde{F} = \frac{8m^2}{\left(r+2m\right)^2}\rd\psi \wedge \rd
r+\frac{4rm\sin\theta} {r+2m}\rd\theta \wedge \rd\phi +
\frac{8m^2\cos\theta}{\left(r+2m\right)^2} \rd\phi \wedge\rd r.
\end{equation}
The fact $F(R,Y) = F(R,*Y)$ implies that charges corresponding to
$Y$ and $*Y$ are the same, energy is self-dual i.e. energy and dual energy are equal.
Unfortunately in the case of Taub-NUT metric, we cannot define the
charge as a integral over a closed surface, since our spacetime
$M$ is nontrivial bundle $M \rightarrow S^{2}$ and there is no
``sphere at infinity''. However, we may project the form
$\widetilde{F}^{\mu\nu}\rd S_{\mu\nu}$ onto the base of the
fibration, integrate it, and get the result, which depends on $r$:
\begin{displaymath}
\frac{1}{16\pi}\int_{S^2} \widetilde{F}^{\mu\nu}\rd
S_{\mu\nu} = \frac{mr}{r+2m}
\longrightarrow\!\!\!\!\!\!\!\!\!\!\!\! \lower.260em\hbox{${}_{r
\to \infty}$} m.
\end{displaymath}
It is not surprising that the ``dual mass'' cannot be defined as a integral
over a finite two-surface.

For $\widetilde{F}$ there exist a potential $\widetilde{A}$
($\widetilde{F} = \rd\widetilde{A}$) equal to
\begin{equation}\label{Taub-NUT_A}
\widetilde{A} = -\frac{4mr}{r+2m}\rd\psi -
\frac{4mr\cos\theta}{r+2m}\rd\phi,
\end{equation}
\begin{equation}\label{Taub-NUT_Avect}
g^{-1}(\widetilde{A}) = -\frac{1}{m}\partial_{\psi}.
\end{equation}
We see that for such a potential $g^{-1}(\widetilde{A})$ is a
Killing vector field of the metric $g$, and therefore
$\widetilde{F}$ is a Papapetrou field (like $F(R,*Q)$ for the
Kerr metric cf. \cite{JJMLKerr}).

It remains to check, how the conformal Killing tensor related
to~$Y$, $*Y$, and~$Y_i$ look like. Let us introduce the following
notation:

\[ K_{\mu\nu}(Y,Z):= \frac12 \left( Y_{\mu}{}^{\kappa}
Z_{\kappa\nu} + Z_{\mu}{}^{\kappa} Y_{\kappa\nu} \right)\, .\]

In coordinate system $(\psi, r, \theta, \phi)$ the easiest way of
writing out these tensors is to treat them as covariant tensors
(with lowered indices). We have the following non-vanishing
components of the symmetric $5\times 5$ matrix $K(\cdot ,\cdot )$:
{\setlength\arraycolsep{2pt}
\begin{eqnarray}
K(Y,Y) & = & -\frac1{1+\frac{2m}{r}}\bigg[4m^4\rd\psi^2 +
8m^4\cos\theta\rd\psi\rd\phi
+ (m+r)^2(r+2m)^2\rd\theta^2 + {}\nonumber\\[10pt]
& & {} + \left(r\sin^2\theta(3m+r)(4m^2 + 3mr +r^2) + 4m^2 \right)
\rd \phi^2 + m^2(1+\frac{2m}{r})^2\rd r^2\bigg],\nonumber
\end{eqnarray}}
{\setlength\arraycolsep{2pt}
\begin{eqnarray}
K(Y_i,Y_j) & = & -\frac1{1+\frac{2m}{r}}\bigg[16m^2\rd\psi^2 +
32m^2\cos\theta\rd\psi\rd\phi
+ 4(r+2m)^2\rd \theta^2 + {}\nonumber\\[10pt]
& & {} + 4\left(r(r+4m)\sin^2\theta  + 4m^2\right)\rd\phi^2 +
\frac{4(r+2m)^2}{r^2}\rd r^2 \bigg]\delta_{ij}\nonumber
\end{eqnarray}}
{\setlength\arraycolsep{2pt}
\begin{eqnarray}
K(Y, Y_1) & = & - \frac1{1+\frac{2m}{r}}\Big[ 8m^3\sin\theta\cos\phi \rd\psi^2 - 2(m+r)(r+2m)^2\sin\theta\cos\phi \rd\theta^2 + \nonumber\\
& & + 2 \Sigma \sin\theta\cos\phi \rd\phi^2 + \frac{2m(r+2m)^2\sin\theta\cos\phi}{r^2}\rd r^2 + \nonumber\\
& & - 2m(r+2m)^2\sin\phi \rd\psi\rd\theta - 2mr(r+4m)\sin\theta\cos\theta\cos\phi \rd\psi \rd\phi + \nonumber\\
& & - 2m(r+2m)^2\cos\theta\sin\phi \rd\theta\rd\phi + \frac{(r+2m)^3\cos\theta\cos\phi}{r}\rd\theta\rd r + \nonumber\\
& & - \frac{(r+2m)^3\sin\theta\sin\phi}{r}\rd\phi\rd r \Big] \nonumber\\
K(Y, Y_2) & = & - \frac1{1+\frac{2m}{r}}\Big[
8m^3\sin\theta\sin\phi \rd\psi^2
- 2(m+r)(r+2m)^2\sin\theta\sin\phi \rd\theta^2 + \nonumber\\
& & + 2\Sigma \sin\theta\sin\phi \rd \phi^2 + \frac{2m(r+2m)^2\sin\theta\sin\phi}{r^2}\rd r^2 \nonumber\\
& & + 2m(r+2m)^2\cos\phi \rd\psi\rd\theta - 2mr(r+4m)\sin\theta\cos\theta\sin\phi \rd\psi\rd\phi \nonumber\\
& & + 2m(r+2m)^2 \cos\theta \cos\phi \rd\theta \rd\phi + \frac{(r+2m)^3\cos\theta\sin\phi}{r} \rd\theta \rd r + \nonumber\\
& & + \frac{(r+2m)^3\sin\theta\cos\phi}{r} \rd\phi \rd r\Big] \nonumber\\
K(Y, Y_3) & = & - \frac1{1+\frac{2m}{r}}\Big[ 8m^3\cos\theta \rd\psi^2 - 2(m+r)(r+2m)^2\cos\theta \rd\theta^2 + \nonumber\\
& & + 2(\Sigma + 2m(r+2m)^2)\cos\theta \rd\phi^2 + \frac{2m(r+2m)^2\cos\theta}{r^2}\rd r^2 + \nonumber\\
& & + 2m(r^2\sin^2\theta + 4mr\sin^2\theta + 4m^2) \rd\psi \rd\phi -
\frac{(r+2m)^3\sin\theta}{r} \rd\theta \rd r \Big] \nonumber\\
K(Y, *Y) & = & - \frac1{1+\frac{2m}{r}}\Big[ 4m^3(m+r) \rd\psi^2 +
m(m+r)(r+2m)^2 \rd \theta^2 +
\nonumber\\
& & + m(m+r)(4m^2 + r(r + 4m)\sin^2\theta )\rd \phi^2 + \nonumber\\
& & + \frac{m(r+2m)^2(m+r)}{r^2}\rd r^2 + 4m^3 (m+r)
\cos \theta \rd \psi \rd \phi \Big] \nonumber\\
K(*Y, Y_1) & = & - \frac1{1+\frac{2m}{r}}\Big[ 8m^2(m+r)\sin\theta\cos\phi \rd\psi^2 - 2m(r+2m)^2\sin\theta\cos\phi \rd\theta^2 + \nonumber\\
& & - 2m(r^2\cos^2\theta + (r+2m)^2)\sin\theta\cos\phi\rd \phi^2 + \frac{2(m+r)(r+2m)^2\sin\theta \cos\phi}{r^2} \rd r^2 + \nonumber\\
& & - 2m(r+2m)^2\sin\phi \rd\psi \rd\theta - 2mr^2 \sin\theta\cos\theta \cos\phi \rd\psi \rd\phi + \nonumber\\
& & - 2m(r+2m)^2\cos\theta\sin\phi \rd\theta \rd\phi + \frac{(r+2m)^3\cos\theta\cos\phi}{r} \rd\theta \rd r + \nonumber\\
& & - \frac{(r+2m)^3\sin\theta\sin\phi}{r} \rd\phi \rd r \Big] \nonumber\\
K(*Y, Y_2) & = & - \frac1{1+\frac{2m}{r}}\Big[ 8m^2(m+r)\sin\theta \sin\phi \rd\psi^2 - 2m(r+2m)^2 \sin\theta \sin\phi \rd \theta^2 + \nonumber\\
& & - 2m(r^2\cos^2\theta + (r+2m)^2) \sin\theta \sin\phi \rd\phi^2 + \frac{2(m+r)(r+2m)^2\sin\theta\sin\phi}{r^2} \rd r^2 + \nonumber\\
& & + 2m(r+2m)^2\cos\phi \rd\psi \rd\theta - 2mr^2\sin\theta\cos\theta \sin\phi \rd\psi \rd\phi + \nonumber\\
& & + 2m(r+2m)^2 \cos\theta \cos\phi \rd\theta \rd\phi + \frac{(r+2m)^3\cos\theta\sin\phi}{r} \rd\theta \rd r + \nonumber\\
& & + \frac{(r+2m)^3\sin\theta\cos\phi}{r} \rd\phi \rd r
\Big] \nonumber\\
K(*Y, Y_3) & = & - \frac1{1+\frac{2m}{r}}\Big[  8m^2(m+r)\cos\theta \rd\psi^2  - 2m(r+2m)^2\cos\theta \rd\theta^2 + \nonumber\\
& & + 2m(4m^2 + 4mr + r^2\sin^2\theta)\cos\theta \rd\phi^2 + \frac{2(m+r)(r+2m)^2\cos\theta}{r^2} \rd r^2 + \nonumber\\
& & + 2m(4m^2 + 4mr + r^2\sin^2\theta) \rd\psi \rd\phi - \frac{(r+2m)^3\sin\theta}{r} \rd\theta \rd r \Big] \nonumber
\end{eqnarray}}
{\setlength\arraycolsep{2pt}
\begin{eqnarray}
K(*Y,*Y) & = & -\frac1{1+\frac{2m}{r}}\bigg[4m^2(m+r)^2\rd\psi^2 +
8m^2(m+r)^2\cos\theta
\rd\psi\rd\phi + {}\nonumber\\[10pt]
& & {} + m^2(4mr\cos^2\theta + 3r^2\cos^2\theta+r^2+4mr+4m^2)
\rd\phi^2+\nonumber\\[10pt]
& & {} + m^2(r+2m)^2\rd\theta^2 + \frac{(m+r)^2(r+2m)^2}{r^2}\rd
r^2\bigg],\nonumber
\end{eqnarray}}
where
$\Sigma := -r^2(r+3m)\sin^2\theta -2m(r^2 + 4mr +2m^2)$.
Matrix $K(\cdot , \cdot)$ contains  $4\times 4$ symmetric sub-matrix of Killing
tensors (corresponding to Yano tensors $Y_i$ and $Y$) with 5
non-vanishing terms. The remaining 5 components are conformal
Killing tensors for Taub-NUT metric.

\section{Conformal rescaling of CYK tensors}\label{conf_resc_ch}
In this section we will be dealing with conformal transformations
and their impact on conformal Yano-Killing tensors. Since most of
the consideration here is independent of the dimension of a
manifold, we will not restrict ourselves to spacetime of dimension
four. We assume that we are dealing with $n$-dimensional manifold
and that this manifold has a metric $g$ (signature of $g$ plays no
role).
\subsection{Basic formulae}
Let
$\Gamma^\alpha{}_{\mu\nu}$ denotes Christoffel symbols of Levi-Civita
connection associated with the metric $g$. We have:
\begin{equation}\label{conn_coef}
    \Gamma^\alpha{}_{\mu\nu} = \frac12 g^{\alpha\beta}(g_{\beta\mu,\nu}
    + g_{\beta\nu,\mu} - g_{\mu\nu,\beta}).
\end{equation}
Let $\tilde{g}$ be conformally rescaled metric, i.e.
$\tilde{g}_{\mu\nu} :=\Omega^2 g_{\mu\nu}$ (and what follows,
$\tilde{g}^{\mu\nu}:=\Omega^{-2} g^{\mu\nu}$), where $\Omega$ is a
certain positive function ($\Omega>0$). We will denote Christoffel
symbols of this metric by $\tilde{\Gamma}^\alpha{}_{\mu\nu}$,
and covariant derivative associated with them by
$\tilde{\nabla}_\mu$. Obviously, for $\tilde{g}_{\mu\nu}$ and
$\tilde{\Gamma}^\alpha{}_{\mu\nu}$ we have formula analogous
to~(\ref{conn_coef}). We have: {\setlength\arraycolsep{2pt}
\begin{eqnarray}\label{conn_coef_conf}
    \tilde{\Gamma}^\alpha{}_{\mu\nu} & = & \frac12 \tilde{g}^{\alpha\beta}
    (\tilde{g}_{\beta\mu,\nu} + \tilde{g}_{\beta\nu,\mu} - \tilde{g}_{\mu\nu,\beta})
    \nonumber\\[6pt]
    & = & \frac12 \Omega^{-2}g^{\alpha\beta}\left((\Omega^2g_{\beta\mu}){}_{,\nu})
    + (\Omega^2g_{\beta\nu}){}_{,\mu} + (\Omega^2g_{\mu\nu}){}_{,\beta}\right)
    \nonumber\\[6pt]
    & = & \Gamma^\alpha{}_{\mu\nu} + \frac12\Omega^{-2}g^{\alpha\beta}
    \left((\Omega^2)_{,\nu}g_{\beta\mu} + (\Omega^2)_{,\mu}g_{\beta\nu}
    + (\Omega^2)_{,\beta}g_{\mu\nu}\right)\nonumber\\[6pt]
    & = & \Gamma^\alpha{}_{\mu\nu} + g^{\alpha\beta} \left(U_{,\nu}g_{\beta\mu}
    + U_{,\mu}g_{\beta\nu} + U_{,\beta}g_{\mu\nu}\right)\nonumber\\[8pt]
    & = & \Gamma^\alpha{}_{\mu\nu} + \delta^\alpha{}_\mu U_{,\nu} + \delta^\alpha{}_{\nu}
    U_{,\mu} - g^{\alpha\beta}U_{,\beta}g_{\mu\nu}\,,
\end{eqnarray}}where $U:=\log\Omega$. Using the formula~(\ref{conn_coef_conf})
and formulas:
\begin{equation}\label{cov_diff_def}
    \nabla_\mu X_{\nu\rho} = X_{\nu\rho,\mu} - X_{\alpha\rho}
    \Gamma^\alpha{}_{\nu\mu} - X_{\nu\alpha}\Gamma^\alpha{}_{\rho\mu}
\end{equation}
and
\begin{equation}\label{cov_diff_def_conf}
    \tilde{\nabla}_\mu X_{\nu\rho} = X_{\nu\rho,\mu} - X_{\alpha\rho}
    \tilde{\Gamma}^\alpha{}_{\nu\mu} - X_{\nu\alpha}\tilde{\Gamma}^\alpha{}_{\rho\mu}
\end{equation}
(which are true for any tensor $X_{\mu\nu}$), we get:
{\setlength\arraycolsep{2pt}
\begin{eqnarray}\label{cov_der_conf_resc}
    \tilde{\nabla}_\mu X_{\nu\rho} & = & \nabla_\mu X_{\nu\rho} - X_{\mu\rho}U_{,\nu}
    - X_{\nu\mu}U_{,\rho} - 2X_{\nu\rho}U_{,\mu} + {}\nonumber\\[8pt]
    & & {}  + g^{\alpha\beta}U_{,\beta}\left( X_{\alpha\rho}g_{\mu\nu}
    + X_{\nu\alpha}g_{\mu\rho}\right).
\end{eqnarray}}

\subsection{Reducing CYK tensors to Yano tensors}
The following theorem has been proved in \cite{JJMLKerr}:
\begin{Theorem}\label{conf_resc_th}
    If $Q_{\mu\nu}$ is a CYK tensor for the metric $g_{\mu\nu}$,
    then $\Omega^3 Q_{\mu\nu}$ is a CYK tensor for the conformally rescaled metric
    $\Omega^2 g_{\mu\nu}$.
\end{Theorem}
The situation which we were dealing here and in \cite{JJMLKerr}
 was the particular one: each considered CYK tensor either was
 Yano tensor or had the dual one being Yano tensor. We can ask
 whether this situation is truly particular. Obviously, if we
 conformally rescale the metric and its CYK tensors (according to
 the Theorem 2 in \cite{JJMLKerr}), then we get CYK tensors which
 in general are not Yano tensors, even if they were Yano before
 the rescaling. However, we can ask if a well chosen conformal
 factor can bring the situation to one we were dealing with in
 \cite{JJMLKerr}. In the case of manifold of dimension different
 than four we cannot define dual CYK tensor. Nevertheless we can
 ask \emph{whether every CYK tensor can be reduced to Yano tensor by a
 properly chosen conformal transformation}.

Let $g_{\mu\nu}$ be a metric of a manifold $M$ and $Q_{\mu\nu}$
its CYK tensor. If $\Omega$ is a positive function, then according
to the Theorem~\ref{conf_resc_th}, $\tilde{Q}_{\mu\nu}:=\Omega^3
Q_{\mu\nu}$ is a CYK tensor of the metric
$\tilde{g}_{\mu\nu}:=\Omega ^2g_{\mu\nu}$. The necessary and
sufficient condition for $\tilde{Q}_{\mu\nu}$ being Yano tensor is
vanishing of $\tilde{\xi}_\mu$ defined  as follows
\begin{equation}\label{div_Q_conf}
    \tilde{\xi}_{\rho}:=\tilde{g}^{\mu\nu}\tilde{\nabla}_{\mu}\tilde{Q}_{\nu\rho}\,.
\end{equation}
>From the formula (cf. Appendix A of \cite{JJMLKerr})
\begin{eqnarray}\label{div_Q_conf_scal}
    \tilde{\xi}_\rho
& = & \Omega\left( \xi_\rho + (n-1)g^{\mu\nu} Q_{\mu\rho}U_{,\nu}\right),
\end{eqnarray} where $\xi_\rho:=g^{\mu\nu}\nabla_{\mu}Q_{\nu\rho}$,
we get that this condition is equivalent to:
\begin{equation}\label{CYK_to_Yano}
    \xi_\rho = (1-n)g^{\mu\nu} Q_{\mu\rho}U_{,\nu}\,,
\end{equation}
where $\xi_\rho:=g^{\mu\nu}Q_{\mu\rho}{}_{;\nu}$ and $U:=\log
\Omega$. We see that CYK tensor $Q_{\mu\nu}$ can be reduced to
Yano tensor if and only if there exist an exact form $\zeta_\mu$
such that
\begin{equation}\label{CYK_to_Yano2}
\xi^\nu=Q^{\mu\nu}\zeta_{\mu}
\end{equation}
(namely $\zeta_\mu = (1-n)U_{,\mu}$). It seems that there are no
reasons for claiming that for every CYK tensor there exist
$\zeta_\mu$ fulfilling the equation~(\ref{CYK_to_Yano2}) and
moreover that this $\zeta_\mu$ is exact (or even closed). In
particular, it is easy to show an example of a CYK tensor for
which there are no globally defined form $\zeta_\mu$.
Let us
consider Minkowski spacetime
($g_{\mu\nu}=\eta_{\mu\nu}=\textrm{diag}(-1,1,1,1)$ in Cartesian
coordinates $x^\mu$).
Let us denote by ${\cal D}$ a {\em dilation vector field}:
\begin{equation}\label{dil_generators}
{\cal D} := x^{\mu}\frac{\partial}{\partial x^{\mu}}\, ,
\end{equation}
and by ${\cal T}_\mu$, ${\cal L}_{\mu\nu}$ generators of Poincare group:
\begin{equation}\label{kvf_generators}
{\cal T}_\mu := \frac{\partial}{\partial x^{\mu}},\quad
{\cal L}_{\mu\nu} := x_{\mu}\frac{\partial}{\partial x^{\nu}} -
x_{\nu}\frac{\partial}{\partial x^{\mu}} \, .
\end{equation}

The CYK tensor $Q={\cal D}\wedge{\cal T}_0$
has its corresponding vector $\xi$ equal to $\frac32\partial_0$. Let us notice
that ${\cal D}$ vanishes in the point $x^\mu=0$, $\mu=0,\ldots,3$.
It means that in this point $Q=0$ and from the
equation~(\ref{CYK_to_Yano2}) we get that for any $\zeta$ the
field $\xi$ is equal to zero which contradict the fact that
$\xi=\frac32\partial_0$. Therefore the
equation~(\ref{CYK_to_Yano2}) cannot be fulfilled everywhere.

We can now check how the situation changes for the case of the Kerr metric:
\begin{equation}\label{Kerr_metric}
g_{\mbox{\tiny\rm Kerr}} = g_{tt} \rd t^2 + 2g_{t\phi} \rd t \rd \phi + g_{rr} \rd r^2
+ g_{\theta\theta} \rd \theta^2 + g_{\phi\phi} \rd \phi^2\, ,
\end{equation}
where
\[
g_{tt} = -1 +{2mr \over \rho ^2},  \quad g_{t\phi} = -{2mra\sin ^2 \theta
\over \rho ^2}, \quad g_{rr} = {\rho ^2 \over \triangle}, \quad
g_{\theta \theta} = \rho^2,
\]
\begin{equation}\label{Kerr_metric_compts}
g_{\phi\phi} =
\sin^2\theta\left(r^2+a^2+{2mra^2\sin^2\theta\over\rho^2}\right),
\end{equation}
with
\begin{equation}\label{Kerr_metric_symbols}
 \rho ^2 = r^2 + a^2 \cos ^2 \theta \quad \textrm{and} \quad
 \triangle = r^2 -2mr +a^2\, ,
\end{equation}
and for the Euclidean Taub-NUT metric given by (\ref{Taub-NUT_metric}).
In spacetimes (\ref{Kerr_metric}-\ref{Kerr_metric_compts}) and (\ref{Taub-NUT_metric})
we have the following CYK tensors:
\begin{equation}\label{Kerr_Q}
Q_{\mbox{\tiny\rm Kerr}}=
r\sin\theta \rd\theta \wedge \left[ \left( r^2+a^2\right)\rd\phi - a\rd t
\right]+ a\cos\theta \rd r\wedge\left(\rd t - a\sin^2\theta \rd\phi \right).
\end{equation}
\begin{equation}\label{Kerr_Qstar}
    \ast Q_{\mbox{\tiny\rm Kerr}}
    =a\cos\theta\sin\theta \rd\theta \wedge \left[ \left(r^2+a^2
    \right) \rd\phi - a\rd t \right]+ r\rd r \wedge \left(a\sin^2\theta \rd\phi -
    \rd t \right)
\end{equation}
in Kerr spacetime (see \cite{JJMLKerr}) and
\begin{equation}\label{Taub-NUT_Ybis}
Y_{\mbox{\tiny\rm NUT}}=2m^2\left(\rd\psi+\cos\theta \rd\phi\right) \wedge
\rd r +r\left(r+m\right)
\left(r+2m\right) \sin\theta \rd\theta \wedge \rd\phi,
\end{equation}
\begin{equation}\label{Taub-NUT_Ystarbis}
\ast Y_{\mbox{\tiny\rm NUT}}=2m\left(m+r\right)\left(\rd\psi+\cos\theta \rd\phi\right)
\wedge \rd r+mr\left(r+2m \right) \sin\theta \rd\theta \wedge \rd\phi.
\end{equation}
for Taub-NUT respectively (cf. \ref{Taub-NUT_Y} and \ref{Taub-NUT_Ystar}).
It turns out that
$*Y_{\mbox{\tiny\rm NUT}}$ defined by the
formula~(\ref{Taub-NUT_Ystar}) can be reduced to Yano tensor by a
properly chosen conformal transformation. For $*Q_{\mbox{\tiny\rm
Kerr}}$ defined by the formula~(\ref{Kerr_Qstar}) we can only find
$U$ satisfying the equation~(\ref{CYK_to_Yano}) which is not
defined on the plane $\theta=\pi / 2$. We will discuss this case
first.

We denoted by $g_{\mbox{\tiny\rm Kerr}}$ the metric tensor defined by the
formulae~(\ref{Kerr_metric})--(\ref{Kerr_metric_symbols}). $*Q_{\mbox{\tiny\rm Kerr}}$
given by~(\ref{Kerr_Qstar}) is a CYK tensor of the
metric $g_{\mbox{\tiny\rm Kerr}}$, but it is not its Yano tensor, since we
have $\chi^\nu:=*Q^{\mu\nu}{}_{;\mu}=3\delta^\nu{}_t$. Rewriting the
equation~(\ref{CYK_to_Yano}) for $*Q$ and~$\chi$ we get:
\begin{equation}\label{Kerr_Qstar_conf_resc}
    \chi^\nu = -3\, {*Q}^{\mu\nu}U_{,\mu}
\end{equation}
(obviously indices in $*Q$ were raised with respect to the metric $g_{\mbox{\tiny\rm Kerr}}$). The only
non-vanishing components of the tensor $*Q^{\mu\nu}$ are the
following:
\[
*Q^{\theta t}=\frac{a^2\sin\theta\cos\theta}{\rho^2}, \quad
*Q^{rt}=\frac{r(r^2+a^2)}{\rho^2},
\]
\[
*Q^{\theta\phi}=\frac{a\cos\theta} {\sin\theta\rho^2} \quad
\textrm{i} \quad *Q^{r\phi}=\frac{ar}{\rho^2}.
\]
Let us write the components of
equation~(\ref{Kerr_Qstar_conf_resc}):
\begin{equation}\label{Kerr_Qstar_conf_resc0}
    1 = -*Q^{\mu t}U_{,\mu} = Q^{t\theta}U_{,\theta}+Q^{tr}U_{,r},
\end{equation}
\begin{equation}\label{Kerr_Qstar_conf_resc1}
    0 = -*Q^{\mu r}U_{,\mu} = Q^{rt}U_{,t}+Q^{r\phi}U_{,\phi},
\end{equation}
\begin{equation}\label{Kerr_Qstar_conf_resc2}
    0 = -*Q^{\mu\theta}U_{,\mu} = Q^{\theta t}U_{,t}+Q^{\theta\phi}U_{,\phi},
\end{equation}
\begin{equation}\label{Kerr_Qstar_conf_resc3}
    0 = -*Q^{\mu\phi}U_{,\mu} = Q^{\phi\theta}U_{,\theta}+Q^{\phi r}U_{,r}.
\end{equation}
>From equations~(\ref{Kerr_Qstar_conf_resc1}) and~(\ref{Kerr_Qstar_conf_resc2}) we get that
\[
-\frac{Q^{r\phi}}{Q^{rt}} U_{,\phi} = U_{,t} =
-\frac{Q^{\theta\phi}}{Q^{\phi t}} U_{,\phi}\,.
\]
If $U_{,\phi}\ne 0$, then we obtain the following contradiction:
\[
\frac{a}{r^2+a^2} = \frac{Q^{r\phi}}{Q^{rt}} =
\frac{Q^{\theta\phi}}{Q^{\phi t}} = \frac1{a\sin\theta}
\]
which implies that $U_{,\phi} = U_{,t} = 0$. Moreover, from
equations~(\ref{Kerr_Qstar_conf_resc0})
and~(\ref{Kerr_Qstar_conf_resc3}) we get
\[
    1 = (Q^{tr} - Q^{t\theta}\frac{Q^{\phi r}}{Q^{\phi\theta}})U_{,r}
    = -r U_{,r}.
\]
Therefore $U=-\log r + f(\theta)$, where $f$ is a certain function
of one variable.
>From the equation~(\ref{Kerr_Qstar_conf_resc3}) we get the following ODE:
\[
    \frac{\rd f}{\rd \theta} = U_{,\theta} = -\frac{Q^{\phi r}}{rQ^{\phi\theta}} =
    \frac{\sin\theta}{\cos\theta}
\]
which maybe easily integrated in closed form:
$f=-\log|\cos\theta| + c_1$, where $c_1$ is a certain
constant. Finally we get that $U=-\log|r\cos\theta|+c_1$ hence
\[
    \Omega = \frac{c_2}{r|\cos\theta|},
\]
where $c_2=e^{c_1}$ is a positive constant. Let us observe that $U$
and $\Omega$ are not defined (or singular) on the plane
$\theta=\frac{\pi}2$. Direct computation shows that tensor $\Omega^3\!
*\!Q$ is indeed Yano tensor of the metric $\Omega^2
g_{\mbox{\tiny\rm Kerr}}$. Tensor $\Omega^3 Q$ is no longer Yano tensor of the
conformally rescaled metric (although obviously it is its CYK
tensor), since we have
$\tilde{\xi}_{\mbox{\tiny\rm Kerr}}=\frac{3a}{c_2}\partial_t+\frac{3}{c_2}
\partial_\phi$ (where $\tilde{\xi}_{\mbox{\tiny\rm Kerr}}$ is defined by (\ref{div_Q_conf})).
The vector field $\tilde{\xi}_{\mbox{\tiny\rm Kerr}}$ is Killing vector field of the metric
$\Omega^2 g_{\mbox{\tiny\rm Kerr}}$. It seems that it is only a coincidence,
since $\Omega^2 g_{\mbox{\tiny\rm Kerr}}$ is no longer an Einstein metric.

Finally, we discuss the case of CYK tensor of Taub-NUT metric. Let
$g_{\mbox{\tiny\rm NUT}}$ denote this metric defined by the
formula~(\ref{Taub-NUT_metric}). Tensor $*Y_{\mbox{\tiny\rm NUT}}$ defined by the
formula~(\ref{Taub-NUT_Ystar}) is its CYK tensor but is not its
Yano tensor, since
\mbox{$\chi^\nu:=*Y^{\mu\nu}{}_{;\mu}=-\frac3{2m}\delta^\nu{}_\psi$}.
In order to find conformal factor $\Omega'$, which reduces $*Y$ to
Yano tensor, let us rewrite the equation~(\ref{CYK_to_Yano}) for
$*Y$, $\chi$ and $U':=\log\Omega'$. We have:
\begin{equation}\label{Taub-NUT_Ystar_conf_resc}
    \chi^\nu = -3*\!Y^{\mu\nu}U'_{,\mu}
\end{equation}
The only non-vanishing components of the tensor $*Y^{\mu\nu}$ are
the following:
\[
*Y^{\psi\theta}=\frac{m\cos\theta}{r(r+2m)\sin\theta}, \quad
*Y^{\psi r}=\frac{m+4}{2m}, \quad
*Y^{\theta\phi}=\frac{m}{r(r+2m)\sin\theta}.
\]
Let us write out the components of the
equation~(\ref{Taub-NUT_Ystar_conf_resc}):
{\setlength\arraycolsep{2pt}
\begin{eqnarray}
1 & = & 2m*\!Y^{\mu\psi}U'_{,\mu} =
2m*\!Y^{\theta\psi}U'_{,\theta}
+ 2m*\!Y^{r\psi}U'_{,r}\,,\label{Taub-NUT_Ystar_conf_resc0}\\
0 & = & 2m*\!Y^{\mu r}U'_{,\mu} = 2m*\!Y^{\psi r}U'_{,\psi}\,,
\label{Taub-NUT_Ystar_conf_resc1}\\
0 & = & 2m*\!Y^{\mu\theta}U'_{,\mu} =
2m*\!Y^{\psi\theta}U'_{,\psi}
+ 2m*\!Y^{\phi\theta}U'_{,\phi}\,,\label{Taub-NUT_Ystar_conf_resc2}\\
0 & = & 2m*\!Y^{\mu\phi}U'_{,\mu} =
2m*\!Y^{\theta\phi}U'_{,\theta}\,,
\label{Taub-NUT_Ystar_conf_resc3}
\end{eqnarray}}Equations~(\ref{Taub-NUT_Ystar_conf_resc1})--(\ref{Taub-NUT_Ystar_conf_resc3})
imply that $U'_{,\psi}=U'_{,\theta}=U'_{,\phi}=0$, that is
$U'=U'(r)$. Using this result and the
equation~(\ref{Taub-NUT_Ystar_conf_resc0}) we get:
\[
    1 = \frac{\rd U'}{\rd r}2m*\!Y^{r\psi} = -(m+r)\frac{\rd U'}{\rd r},
\]
that is $U'=-\log(m+r)+c_3$, where $c_3$ is a certain constant.
Finally:
\[
    \Omega' = \frac{c_4}{m+r},
\]
where $c_4=e^{c_3}$ is a positive constant. Again direct
computation shows that $\Omega'^3 *\!Y$ is a Yano tensor of
the metric $\Omega'^2 g_{\mbox{\tiny\rm NUT}}$. The tensor $\Omega'^3 Y$
(where $Y$ is defined by the formula~(\ref{Taub-NUT_Y})) is a CYK
tensor of the conformally rescaled metric, for which we have:
$\tilde{\xi}_{\mbox{\tiny\rm NUT}}=\frac{3}{2c_4}\partial_\psi$. Although $\Omega'^2
g_{\mbox{\tiny\rm NUT}}$ is no longer an Einstein metric, a vector $\tilde{\xi}_{\mbox{\tiny\rm NUT}}$
is its Killing vector.

Considerations in this Section show that there is no unique answer to the
question if we can reduce CYK to Yano via conformal rescaling,
for Kerr the answer is negative but for Euclidean Taub-NUT is positive.


\begin{thebibliography}{66}

 \bibitem{Berg} G. Bergqvist, P. Lankinen, \emph{Unique characterization
 of the Bel--Robinson tensor}, Classical and Quantum Gravity 21,
 (2004) 3489--3503; G. Bergqvist, I. Eriksson and J.M.M. Senovilla,
 \emph{New electromagnetic conservation laws}, Classical and Quantum
 Gravity 20, (2003) 2663--68
\bibitem{Sen} M.A.G. Bonilla and J.M.M. Senovilla,
 \emph{Some Properties of the Bel and Bel--Robinson Tensors}
 General Relativity and Gravitation 29, (1997) 91--116



\bibitem{Ch-Kl} D. Christodoulou and S. Klainerman, \emph{Asymptotic
Properties of Linear Field Equations in Minkowski Space},
Communications on Pure and Applied Mathematics 43, (1990) 137--199
\bibitem{Douglas} S.R. Douglas, \emph{Letter: Review of the Definitions
of the Bel and Bel-Robinson Tensors}, General Relativity and
Gravitation 35, (2003) 1691--97
\bibitem{Gib-Riet-vHolt} G.W. Gibbons,  R.H. Rietdijk and J.W. van Holten,
\emph{SUSY in the sky}, Nucl. Phys.~B 404, (1993) 42--64,
{\arxiv{hep-th/9303112}}
\bibitem{Glass-Naber} E.N. Glass and M.G. Naber, \emph{Gravitational mass
anomaly}, Journal of Mathematical Physics 35, (1994) 4178--83
\bibitem{Goldberg1} J.N. Goldberg,  \emph{Conserved quantities
at spatial and null infinity: The Penrose potential}, Phys. Rev. D
41, (1990) 410--417
\bibitem{vHolt} J.W. van Holten, \emph{Phys. Lett.} B \textbf{342}, (1995) 47-52
\bibitem{JJspin2} J. Jezierski, \emph{The Relation between Metric and
Spin--2 Formulations of Linearized Einstein Theory},
 Gen. Rel. and Grav. 27, (1995) 821--843, \arxiv{gr-qc/9411066}
\bibitem{kerrnut} J. Jezierski,
 {\it  Conformal Yano--Killing tensors and asymptotic CYK tensors for
 the Schwarzschild metric},
 Classical and Quantum Gravity 14, (1997) 1679--1688, \arxiv{hep-th/9411074}

\bibitem{JJschwarzl} J. Jezierski, {\it Gauge-invariant formulation of the
linearized Einstein equations  on the Schwarzschild background},
in Current Topics in Math. Cosmology, eds M. Rainer and H-J
Schmidt, World Scientific (1998); {\it Energy and angular momentum
of the weak gravitational waves on the Schwarzschild background --
quasilocal gauge-invariant formulation}, General Relativity and
Gravitation 31,  (1999) 1855--1890, \arxiv{gr-qc/9801068}



\bibitem{cykem} J. Jezierski, {\it CYK Tensors, Maxwell Field
   and Conserved Quantities for Spin-2 Field},
   Classical and Quantum Gravity {19}, (2002) 4405--4429, \arxiv{gr-qc/0211039}

\bibitem{JJMLKerr} J. Jezierski, M. {\L}ukasik,
\emph{Conformal Yano-Killing tensor for the Kerr metric and
conserved quantities},
Class. Quantum Grav. 23, (2006) 2895--2918, \arxiv{gr-qc/0510058}

\bibitem{Moroianu} A. Moroianu, U. Semmelmann,
\emph{Twistor forms on K\"ahler manifolds}, Ann. Scuola Norm. Sup.
Pisa Cl. Sci. (4) II (2003), 823--845, \arxiv{math.DG/0204322}






\bibitem{Pen1} R. Penrose, \emph{Quasi-local Mass and Angular Momentum in
General Relativity}, Proc. Roy. Soc. Lond. A381, (1982) 53--62
\bibitem{Pen-Rin} R. Penrose and W. Rindler, {\it Spinors and Space-time},
Cambridge University Press, Vol. 2, p. 396 (Cambridge 1986)


\bibitem{Semmelmann} U. Semmelmann, \emph{Conformal Killing forms on
Riemannian manifolds}, Mathematische Zeitschrift 245, (2003)
503--527

\bibitem{Stepanow} S.E. Stepanov, \emph{The Vector Space of Conformal
 Killing Forms on a Riemannian Manifold},
 Journal of Mathematical Sciences 110, (2002) 2892--2906;
\emph{On conformal Killing 2-form of the electromagnetic field},
Journal of Geometry and Physics 33, 
(2000) 191--209

\bibitem{SKM} H. Stephani et al., \emph{Exact solutions of Einstein's
 field equations}, 2nd ed., University Press (Cambridge 2003)


\bibitem{Tachibana} S. Tachibana,
\emph{On conformal Killing tensor in a Riemannian space}, Tohoku
Math. J. (2) 21, (1969) 56--64; T. Kashiwada,  \emph{On conformal
Killing tensor}, Natur. Sci. Rep. Ochanomizu Univ. 19, (1968)
67--74; S. Tachibana  and T. Kashiwada,  \emph{On the
integrability of Killing--Yano's equation}, J. Math. Soc. Japan
21, (1969) 259--265


\bibitem{Yano} K. Yano, \emph{Some remarks on tensor fields and curvature},
Ann. Math. {55}, (1952) 328--347

\end{thebibliography}
\end{document}